# Phonons and Thermodynamics of LiMPO$_4$ (M=Mn, Fe)

Prabhatasree Goel[1], M. K. Gupta[1], R. Mittal[1], S. Rols[2], S. J. Patwe[3], S. N. Achary[3], A. K. Tyagi[3] and S. L. Chaplot[1]

[1]*Solid State Physics Division, Bhabha Atomic Research Centre, Mumbai, 400085, India*
[2]*Institut Laue-Langevin, BP 156, 38042 Grenoble Cedex 9, France*
[3]*Chemistry Division, Bhabha Atomic Research Centre, Mumbai,400085, India*

Lithium transition metal phospho-olivines are useful electrode materials, owing to their stability, high safety, low cost and cyclability. We report phonon studies using neutron inelastic scattering experiments, ab-initio density functional theory calculations and potential model calculations on LiMPO$_4$ (M=Mn, Fe) at ambient and high temperature to understand the microscopic picture of Li sub-lattice. The experiments are in good agreement with calculations. The lattice dynamics calculations indicate instability of a zone-centre as well as zone-boundary modes along (100) at volume corresponding to high temperature. The unstable phonon modes show mainly large vibration of Li atoms in the x-z plane of the orthorhombic structure (space group Pbnm). Molecular dynamics simulations with increasing temperature indicate large mean square displacement of Li as compared to other constituent atoms. The computed pair-correlations between various atom pairs show that there is local disorder occurring in the lithium sub-lattice with increasing temperature, while other pairs show minimal changes. The results find the two compounds to be thermally stable up to high temperatures, which is a desirable trait for its battery applications.

## I. INTRODUCTION

Ever increasing energy demands has always promoted research for the elusive perfect battery material, thus paving way for a safer, secure, sustainable energy alternative[1,2,3]. Lithium transition metal phospho-olivines[4,5,6] have emerged as one of the leading contenders in this race. Since the time they were proposed by Padhi[4,5] et al, they have been actively considered as a potential and viable alternative to the conventional cathode materials like $LiCoO_2$. $LiMPO_4$ have large theoretical capacity[7] and a large operating voltage[6,8]. The importance is further accentuated by their stability[9,10], high safety, low cost, low toxicity and cyclability. The electrochemical[11-14] properties of these materials determine Li intercalation and deintercalation, vital to their functioning as useful electrode materials. They in turn are determined by their basic structural and thermodynamic properties. Despite enormous works pertaining to battery research,[11-17] many aspects of their basic vibrational and other physical properties have not been copiously studied. Studies on delithiated[18-21] $LiMPO_4$, solid solutions have been done to explore methods to increase conductivity in these compounds[18-26]. There have been several first principle studies of their electronic properties[26-31], Li diffusion[32-37], easy direction of lithium diffusion[38,39] and some theoretical studies on their vibrational properties and a few Raman, infrared studies on zone centre phonon modes[40-45].

At ambient conditions, these compounds crystallize in olivine type orthorhombic Pnma[46] space group analogous to mineral Triphylite structure. The magnetic structure with spin orientation of the M atom is given in Fig. 1. The structure comprises of discrete $PO_4$ tetrahdera and highly distorted oxygen octahedra about lithium and transition metal ion, M. The $PO_4$ tetrahdra are irregular, with two significantly different sets of O-O distances. The $PO_4$ tetrahedra form a rigid framework.

A rich collection of studies on these compounds relating to its technological applications have been available in literature. There have been studies to determine the preferable directions of Li diffusion in these compounds using Mossbauer as well as first principles method. The chemical diffusion coefficient of Li has been studied using galvanostatic intermittent titration technique[47-49]. Atomistic studies on the role of defects in Li transport have been studied[50]. Effect of non-stoichiometry and doping on Li diffusion[51,52] has also been extensively studied. Doping[24] or vacancies[28] in the lattice are found to increase the conductivity of these compounds.

Our interest in this study has been to understand the dynamics of these compounds and plausible role of phonons in triggering the lithium ion movement. In this paper we have studied phonons in the entire Brillouin zone, softening of phonons with increasing volume, diffusion of lithium and microscopic picture of the disturbances occurring in the lattice with increasing temperature. Here we report the inelastic neutron scattering measurements of the phonon density of states and its calculation using shell model as well as *ab-initio* approach. The model calculations are in good agreement with the inelastic neutron scattering data and *ab-initio* calculations. This allowed testing of our model, which is used to explore the possibility of Li diffusion using molecular dynamics simulations. Section II and III provide details on the inelastic neutron scattering investigations and the computations respectively. Section IV discusses the various results obtained along with their significance. The conclusions obtained from our work are given in Section V.

## II. EXPERIMENTAL

Polycrystalline samples of LiMnPO$_4$ and LiFePO$_4$ were synthesized by the solid state reaction of appropriate reactants. LiMnPO$_4$ was synthesized from the stoichiometric ratio of NH$_4$MnPO$_4$.H$_2$O and Li$_2$CO$_3$ in argon atmosphere at 750 °C. NH$_4$MnPO$_4$.H$_2$O was prepared by precipitation from MnCl$_2$ solutions using aqueous ammonia. LiFePO$_4$ was synthesized by the reaction between Li$_2$CO$_3$, Fe$_2$O$_3$ and NH$_4$H$_2$PO$_4$ and glucose at 700 °C in flowing N$_2$ atmosphere. Glucose was used as reagent for reducing Fe$^{3+}$ to Fe$^{2+}$.

The inelastic neutron scattering measurements of the phonon density of states in the polycrystalline samples of LiMPO$_4$ (M=Mn,Fe) have been performed using the IN4C time of flight spectrometer at the Institut Laue-Langevin (ILL), in Grenoble, France. The measurements at 300 K were performed in the neutron-energy gain mode with incident neutron energy of 14.2 meV (2.4 Å). The detector bank covered scattering angles from 10° to 110°. The signal is corrected for the contributions from the empty cell and suitably averaged over the angular range using the available software package at ILL. A standard vanadium sample was used to calibrate the detectors.

In the incoherent[53] one-phonon approximation the measured scattering function $S(Q,E)$, as observed in the neutron experiments, is related to the neutron cross-section weighted phonon density of states as follows:

$$g^{(n)}(E) = A \left\langle \frac{e^{2W_k(Q)}}{Q^2} \frac{E}{n(E,T) + \frac{1}{2} \pm \frac{1}{2}} S(Q,E) \right\rangle \qquad (1)$$

$$g^n(E) = B \sum_k \{\frac{4\pi b_k^2}{m_k}\} g_k(E) \qquad (2)$$

where the + or − signs correspond to energy loss or gain of the neutrons, respectively, and $n(E,T) = [\exp(E/k_B T) - 1]^{-1}$. A and B are normalization constants and $b_k$, $m_k$, and $g_k(E)$ are, respectively, the neutron scattering length, mass, and partial density of states of the $k^{th}$ atom in the unit cell. The quantity within < ---- > represents an appropriate average over all the neutron momentum transfer (Q) values at a given energy. 2W(Q) is the Debye-Waller factor. The weighting factors $\frac{4\pi b_k^2}{m_k}$ for each atom type in the units of barns/amu are: Li: 0.196; Fe: 0.208 ; Mn: 0.039; P: 0.107 and O: 0.265 calculated from the neutron scattering lengths.[54]

## III. COMPUTATIONAL DETAILS

The lattice dynamics calculations have been performed using both empirical potential[55-57] as well as *ab-initio* methods. Density functional theory (DFT) has been shown to describe the structural and lattice dynamical properties of material using pseudopotentials and plane wave basis sets. In the frame work of density functional perturbation theory (DFPT)[58] it is possible to calculate phonon frequencies, dielectric constants, and other properties. We have used the Vienna ab initio simulation package (VASP)[59,60] along with the PHONON package for the phonon calculations. The volume dependence of zone centre and zone boundary phonon modes have been calculated using density functional perturbation theory implemented in VASP. The plane wave pseudo-potential has been used with maximum plane wave energy cutoff of 500 eV for both the compound. The integrations over the Brillouin zone as been performed on a 4×7×9 grid of k-points generated by Monkhorst[61] pack method. The generalized gradient approximation (GGA) exchange correlation given by Perdew Burke and Ernzerhof [62,63] with projected–augmented wave method has been used. The primitive unit cell of $LiMPO_4$ (M=Fe, Mn) has 28 atoms in a unit cell. The compounds have Fe/Mn magnetic atoms. The magnetic structure of $LiMPO_4$ is same as that of primitive unit cell. The magnetic structure as shown in Fig.1 is used while optimizing the structure as well as calculating the force constants. The self interaction energy correction for the d electrons has been taken care by using LSDA+U approach as introduced by Dudarev[64] et al. The value of the onsite interaction term U=4.3 eV[33] for d electrons for Fe and Mn has been taken from previous studies on $LiMPO_4$ (M=Mn, Fe). For the phonon calculation, we have used a 1×2×2 super cell of the primitive unit cell. A total 112 atoms are contained in the super-cell. The convergence criteria for the total energy and ionic forces were set to $10^{-8}$ eV and $10^{-5}$ eV Å$^{-1}$, respectively. Total energies and inter-atomic forces were calculated for the 16 structures resulting from individual displacements of the three symmetry inequivalent atoms along the three cartesian directions (±x, ±y and ±z). Phonon frequencies were extracted from subsequent calculations using the PHONON software.[65]

Our empirical calculations have been carried using interatomic potentials consisting of Coulomb and short-range Born-Mayer type interaction terms. The form of the interatomic potential used in our model is given by the following expression:

$$V(r) = \frac{e^2}{4\pi\varepsilon_0} \frac{Z(k)Z(k')}{r} + a \exp\left(\frac{-br}{R(k)+R(k')}\right) - \frac{C}{r^6}$$

Where, a=1822eV and b=12.364 are empirical constants. We have successfully used this set of parameters in the lattice dynamical calculations of several complex solids. The term $C_{ij}$ =100 eVÅ$^6$ accounts for the van der Waals interaction between O-O pairs. The effective charge Z(k) and radii R(k) parameters used in our calculations are Z(Li)= 1.0 , Z(Fe/Mn)= 1.869  Z(P)= -3.581.6, Z(O)= -1.6125 , R(Li)= 1. 35 Å,  R(Fe)= 1.88 Å, R(Mn)= 1.91 Å ,  R(P)=  1.20,and R(O)= 1.65 Å. The polarizability of the oxygen atoms is introduced in the framework of the shell model[57]. The shell charge and shell core force constants for oxygen atoms are −2.00 and 60 eV/Å$^2$ respectively. The parameters of the potentials satisfy the conditions of static and dynamic equilibrium. The calculations have been carried out using the current version of DISPR.[66]

Classical molecular dynamics calculations have been carried out to understand the diffusion of lithium ions with increasing temperature. The interatomic potential parameters are the same as those obtained from our lattice dynamics simulations except that the oxygen polarizability was not included. The evolution of the super cell consisting of 7056 ions (4**a** × 7**b** × 9**c**) has been carried out in the NPT ensemble. The mean-square displacements of atoms and pair correlation between pairs of atoms have been calculated using the simulations.

## IV. RESULTS AND DISCUSSION

### A. Phonon Density of States

The measured neutron inelastic scattering spectra at room temperature with incident neutron energy of 14.2 meV are shown in Fig.2. The geometrical constraints of the instruments have allowed collecting the data from about 1 to 7 Å$^{-1}$. We find that for data summed over 1 to 7 Å$^{-1}$ the phonon spectra of both the LiMPO$_4$ (M=Mn, Fe) consist of a strong peak around 5 meV. The Raman and infrared measurements do not indicate zone-centre optic modes at such low energies. The Fe and Mn compounds show antiferromagnetic ordering below 52 and 35 K respectively. In order to understand the origin of the peak, experimental data were grouped in two different regimes namely, high Q (4 to 7 Å$^{-1}$) and low Q (1 to 4 Å$^{-1}$) respectively. We find that for high Q data the intensity of the low energy excitations around 5 meV decreased strongly, while low Q data indicate substantial increase in intensity of the same peak.

Further S(Q,E) plots for LiMPO$_4$ (M=Mn, Fe) are shown in Fig. 3. As mentioned above both the compounds have antiferromagnetic ordering below 50 K. Strong intensity for excitations at 300 K below 10 meV is due to paramagnetic scattering of the compounds. The excitations are more intense at low Q due to magnetic form factor. The higher energy excitations are due to phonons, as their intensities increase with increasing Q.

Fig. 4 gives the neutron weighted density of states of the powder samples in LiFePO$_4$ and LiMnPO$_4$ in comparison with the calculated model potential results and ab-initio calculations. The multiphonon spectra have been calculated using the Sjolander[53] formalism and subtracted from the experimental data. The phonon spectrum extends up to 150 meV. The experimental spectrum consists of several peaks at 25, 55, 75 and 120 meV. No modes are observed for energies between 85 and 105 meV. The general characteristics of the experimental features are well reproduced by the calculations. The measurements are found to be in agreement with our model as well as *ab-initio* calculations. The parital contribution of the constituent atoms to the total phonon density of states in the two olivines computed using both model and *ab-initio* calculations is shown in Fig. 5. We find that Fe/Mn ions contribute largely below 40 meV, while Li being lighter contributes up to 75 meV. The density found beyond 105 meV is only due to the stretching modes of the PO$_4$ polyhedra. Further the comparison of calculated partial density of states (Fig. 5) of various atoms from both the approaches indicates slight difference in the phonon spectra for lithium atoms above 40 meV.

## B. Behavior of Phonons with Volume

The phonon dispersion has been calculated along the high symmetry directions using potential model as well as *ab-initio* method. In general the calculated nature of dispersion relation is nearly same from both the methods, as also reflected in the calculated density of states (Figs. 4 and 5). Fig. 6 shows the calculations of phonon dispersion relations using the *ab-initio* method. The LO-TO (longitudinal optic and transverse optic) splitting of the modes has also been included while plotting the phonon dispersion relation. The phonon frequencies are usually expected to soften with increase of volume. Our previous studies on $Li_2O$ show that the zone-boundary transverse-acoustic phonon mode along [110] softens at volume corresponding to the superionic transition in $Li_2O$[67]. The phonon frequencies of $LiMPO_4$ (M=Mn,Fe) have been calculated (Fig. 7) using *ab-initio* method as a function of volume. The phonon frequencies along all the three high symmetry direction are found to soften with increase of volume. However, the softening is found to be very large for one of the zone-centre (ZC) and zone-boundary (ZB) modes along [100] direction. The change in the phonon frequency with increasing volume (Fig. 8) has been plotted for these ZC and ZB modes. We find that in both the compounds the zone centre optic mode softens first, followed in quick succession by the zone-boundary mode with increasing volume. The thermal expansion behaviour as calculated from molecular dynamics calculations (as discussed below in Section IVB) indicates that $LiFePO_4$ exhibits phonon softening at computed unit cell volume corresponding to 1500 K while similar phenomena occurs at 900 K in $LiMnPO_4$. This region is hitherto defined by us as dynamically unstable regime. The detailed studies of the diffusion of Li in the two compounds are discussed in section IV D below.

For qualitative understanding of the atomic displacement in these unstable modes, we have plotted eigen vectors of both these modes (Fig. 9). In case of ZC mode at ambient volume the displacements of the Li atoms is maximum, while the amplitudes of other atoms are less but not negligible. In the high temperature regime, the amplitude of Li atoms has increased significantly. The Fe atoms are at rest, while the amplitude of P atoms decreased slightly and O atoms do not show any change. The detailed analysis of the direction of lithium displacement in the ZC mode indicates that the net movement of lithium at ambient volume is along face diagonal in the x-y plane but in the high temperature regime the net displacement of Li atoms is in the x-z plane with a higher component along x direction. This suggests that x-z direction is favorable for lithium movement.

For ZB mode (Fig. 9), all the atoms have finite displacement in the ambient regime, while in the high temperature regime, the amplitude of Fe and P atoms have reduced substantially and there is a large increase in amplitude of Li atoms. It is interesting to note that Li atoms at the corners of the unit cell do not show any substantial change in amplitude. This behaviour is different as compared to the ZC mode where amplitude of all Li atoms increased simultaneously. As far as O atoms are concerned, the amplitudes do not change. Careful analysis of the eigen vector of ZB mode also indicates that, initially at ambient volume the net direction of the movement of the Li atoms lie in the x-y plane but in the high temperature regime, the net movement is along the x-z plane, with a larger component along the x-direction.

We find that $LiFePO_4$ shows softening (Fig. 8) at higher volume in comparison with the Mn counterpart. The percentage change in volume for initiation of phonon instability in $LiMnPO_4$ is much lesser as compared to $LiFePO_4$. This may point to the fact that the onset of increased mean-squared amplitude (MSD) of lithium, in $LiMnPO_4$ might occur at a lower temperature. The phonon instabilities observed at higher volume may correspond to higher lithium MSD compared to the ambient volume. The polarization analysis (Fig. 9) of the eigen vectors also suggests the most probable direction for the initiation of lithium movement is along x-direction. The main interest in these compounds stems from their use as battery materials. Lithium intercalation and subsequent delithiation are the main processes by which energy is transferred during its use as battery material. We have so far tried to unveil the role of phonons in the initiation of lithium movement crucial for the use of these materials as battery material. Under suitable conditions[24,28,51,52], such as doping in these materials, there is a possibility that lithium ions may move along these possible directions. This may result in an enhancement of Li conductivity.

## C. Thermal Expansion, Diffusion Coefficient and Pair Correlation Function

Molecular dynamics simulations have been carried out using a supercell of 7056 atoms to study its evolution with increasing temperature. The system has been allowed to evolve for 80 picoseconds at each temperature. The thermal expansion of $LiMPO_4$ (M=Mn,Fe) has been calculated using model potentials in the quasiharmonic approximation and molecular dynamics simulations. The results have been compared with available experimental data[20,40] in Fig. 10. There is a slight deviation between the calculated and experimental thermal expansion for $LiFePO_4$. The quasiharmonic calculations agree very well with the molecular dynamics simulations, suggesting that there are not very explicit temperature effects in the lattice with increasing temperature.

At 300 K, the galvanostatic and potentiostatic[47-49] intermittent titration of the most diffusing member has yielded values between $10^{-14}$ and $10^{-18}$ $m^2/s$ for diffusion of lithium ($D_{Li}$) ions in solid solution of $Li_xFePO_4$ and $Li_{1-x}FePO_4$ (X < 0.02), respectively. GITT, electrochemical impedance spectroscopy (EIS) and cyclic voltammetry (CV) have been carried out to obtain $D_{Li}$[47,48,49] in thin films of $LiFePO_4$, these values are also in the order of $10^{-14}$ $cm^2/s$. Self diffusion coefficient of $Li^+$ at 300 K measured using muon-spin relaxation[35] is ~ $10^{-14}$ $m^2/s$ in $LiFePO_4$. Experimental visualization[38] of lithium diffusion has been reported in delithiated $Li_{0.6}FePO_4$. The diffusion of the constituent atoms in $LiMPO_4$ (M=Mn,Fe) was monitored in our molecular dynamics simulations. As expected lithium exhibits maximum mean square displacement as seen in Fig. 11 in both the compounds, but we did not notice any perceivable diffusion of lithium from our results, which indicates that diffusion coefficient is well below $10^{-10}$ $m^2/s$, beyond the scope of our molecular dynamics simulations. This concludes that these compounds in the pure form cannot be considered to belong to the family of superionic conductors[67,68]. The coefficient of diffusion of lithium in these compounds is reported to be around $10^{-12}$ $m^2/s$, about three orders lower than that of a liquid ($10^{-9}$ $m^2/s$), which is the typical value for a superionic conductor. The various experimentally reported values of $D_{Li}$ in the literature are found to be several orders of magnitude lesser than reported calculated values. The experimentally measured $D_{Li}$ falls in the range of $10^{-17}$ to $10^{-11}$ $m^2/s$, which is the range suitable for a cathode material, hence finds extensive use in rechargeable batteries.

The pair correlations between two atoms indicate the probability of finding a given type of atom with respect to the atom considered. Fig. 12(a) and (b) show the pair correlation between various ion pairs in both the compounds at different temperatures. The correlation between Li ions shows some noticeable changes with increasing temperatures; at 1100 K there are some subtle changes in the atomic arrangement of lithium .This shows that with increase in temperature, there are readjustments occurring continuously in the Li-sub-lattice. In case of correlations between Fe-P, Li-P and Li-Fe there are very small changes indicating that the positions of these atoms have not changed with respect to each other. There are some small changes in the correlations between Fe-O, P-P, Li-O, Fe-Fe and O-O. The P-O correlations are virtually undisturbed with increase in temperature, suggesting that $PO_4$ unit remains rigid, but there are some discernable changes in the orientations of the $PO_4$ polyhedra, particularly with respect to Li-sublattice. The correlations between Li-O changes discernable changes at 1100 K as against the room temperature results. The behavior of the two olivines is similar with increase in temperature. Nevertheless the disturbance in the lattice appears to be very little and does not suggest large scale diffusion in case of both the compounds. The presence of well defined correlation peaks at higher temperature, of the various pair points out to the fact that the compounds are stable and increased disturbances in lithium sub-lattice is a local effect and does not bring any drastic change to the overall lattice.

## V. CONCLUSIONS

This work reports the study of the vibrational and thermodynamics properties of battery materials, $LiFePO_4$ and $LiMnPO_4$. A combination of inelastic neutron scattering, model lattice dynamics, molecular dynamics simulations and first principles calculations has been successfully used to understand the phonon dynamics and high temperature behavior in these compounds. The measured phonon density of states is in excellent agreement with our theoretical calculations. Our molecular dynamics simulations suggest that diffusion coefficient of lithium is less than $10^{-10}$ $m^2/s$ in pure compounds, which is in agreement with reported results. Although lithium has maximum mean square displacement, it does not show any macroscopic diffusion. The pair correlation also suggest that there are subtle changes and disturbances occurring in the Li sub lattice, while $PO_4$ remains structurally unaffected even up to the highest temperatures studied. The phonon density of states calculated using first principles at volume corresponding to high temperature in $LiFePO_4$ and $LiMnPO_4$ suggests that there is a significant instability initiating in the Li sub-lattice. Analysis of the eigen vectors of these modes suggest that if conducive conditions for Li movement from its lattice positions are available, the net displacement of Li atoms is in the x-z plane with a higher component along x direction. Our calculations find the compounds to be thermally stable beyond 1000 K. Our exhaustive studies on their vibrational and thermodynamical properties will be useful in improving the utilization of $LiMPO_4$ (M=Mn,Fe) in numerous technological applications.

FIG. 1(Color online) Magnetic structure[46] of LiMPO$_4$ (M=Mn, Fe) (orthorhombic Pnma space group) derived from *xcrysden* software at T = 0 K. The arrow on the Fe atom indicates the magnetic moment direction. Key; Li: Red spheres, M=Mn or Fe: Yellow spheres, P: Green spheres, O: Blue spheres.

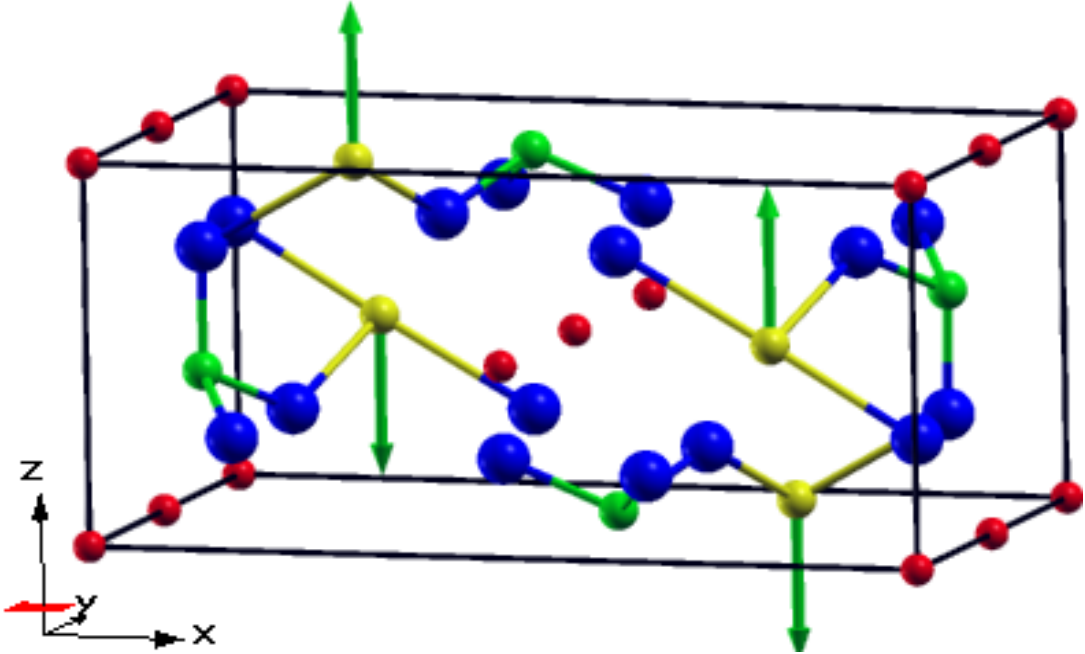

FIG. 2 (Color online) The experimental neutron inelastic scattering spectra of LiMPO$_4$ (M=Mn, Fe) at 300 K. For better visibility the low Q and high Q spectra are shifted along the y-axis by 0.005 meV$^{-1}$ and 0.01 meV$^{-1}$ respectively.

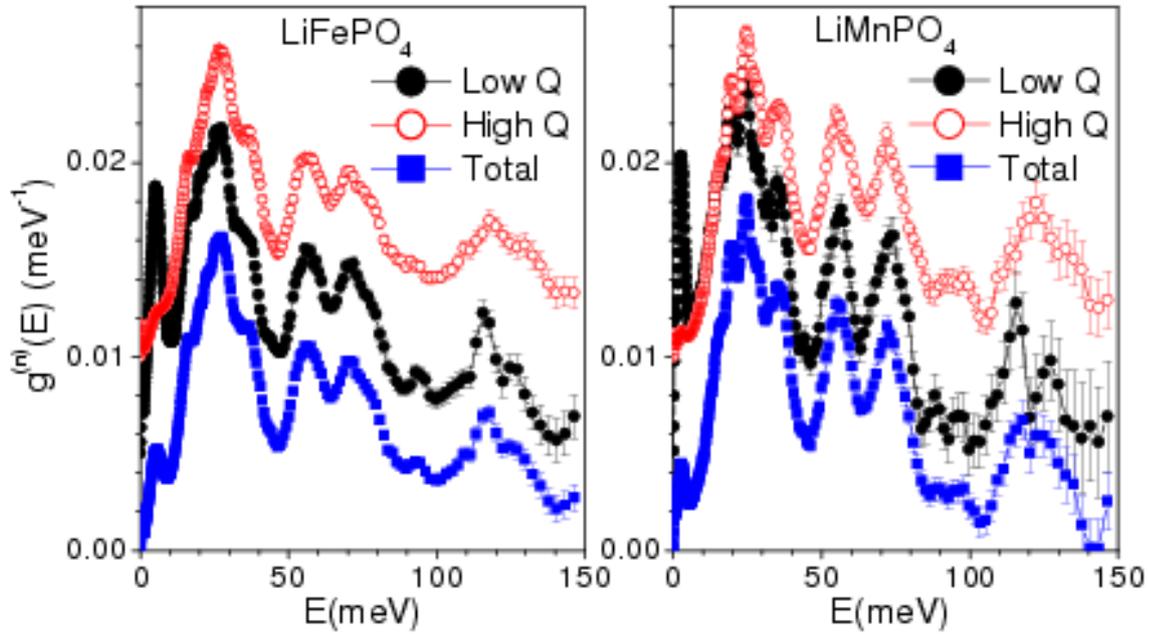

FIG. 3 (Color online) The experimental S(Q,E) plots for LiMPO$_4$ (M=Mn, Fe) at 300 K.

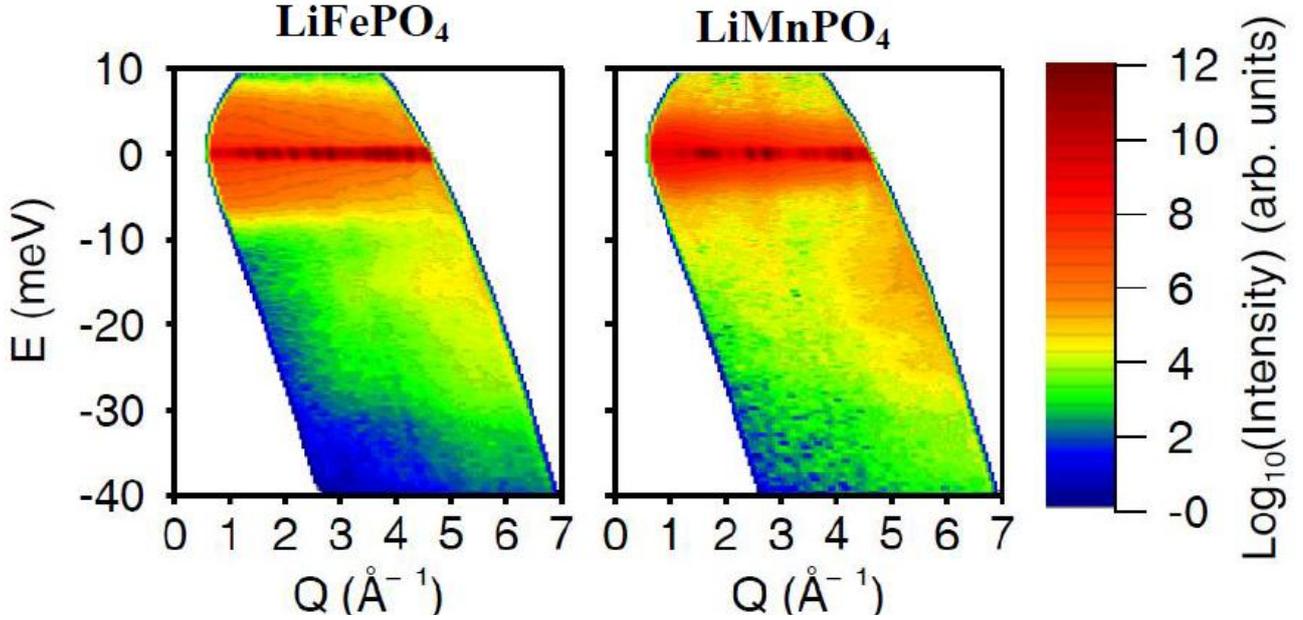

FIG. 4. (Color online) The comparison of the calculated and experimental neutron inelastic scattering spectra for LiMPO$_4$ (M=Mn, Fe) at ambient pressure at 300 K. The *ab-initio* as well as potential model calculations are carried out at 0 K. The multi-phonon contribution calculated using the Sjolander formalism [53] has been subtracted from the experimental data. The experimental spectra comprises of magnetic and phonon contribution, while computed results pertain to phonon contribution alone. The calculated spectra have been convoluted with a Gaussian of FWHM of 10% of the energy transfer in order to describe the effect of energy resolution in the experiment.

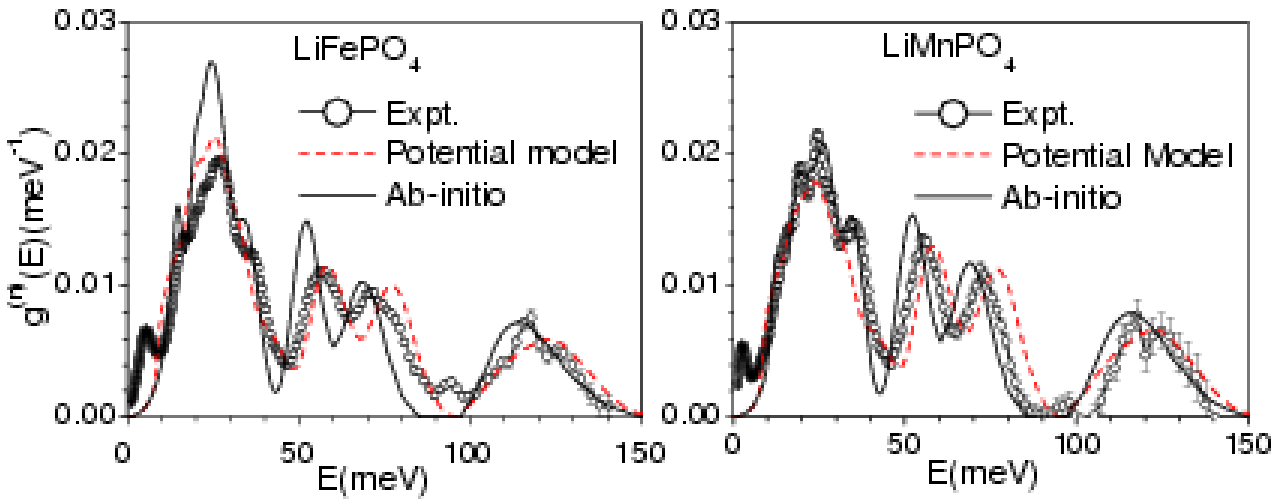

FIG. 5 (Color online) The calculated partial densities of states in LiMPO$_4$ (M=Mn,Fe). The solid and dashed lines correspond to the calculations carried out using *ab-initio* and potential model.

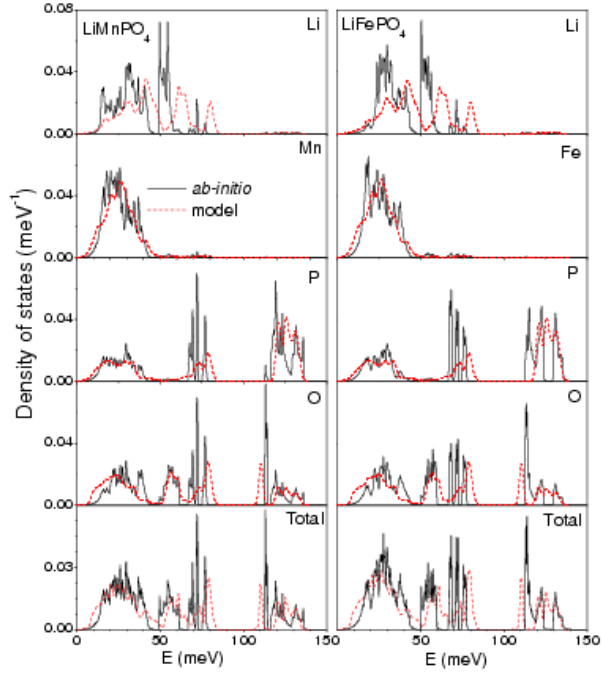

FIG. 6. The Calculated phonon dispersion for LiMPO$_4$ (M=Mn, Fe) from *ab-initio* density functional theory under generalized gradient approximation (GGA-DFT).

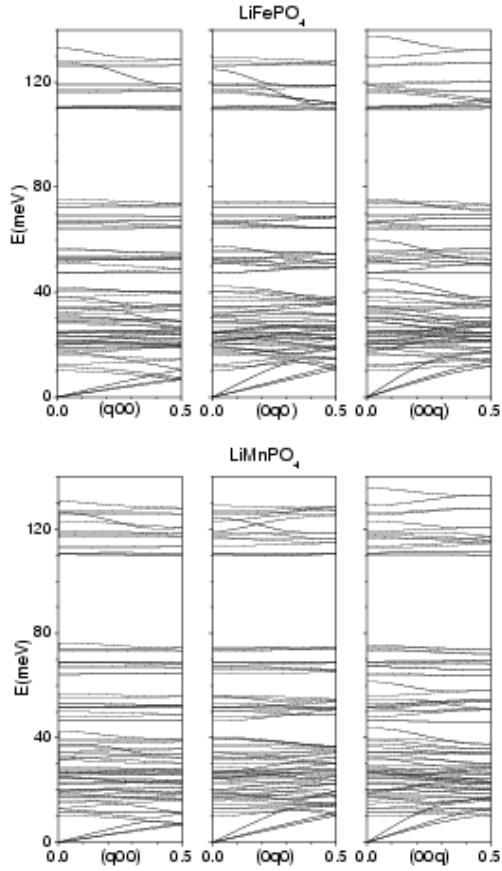

FIG. 7. (Color online) The low-energy part of the phonon dispersion relation from *ab-initio* density functional theory under generalized gradient approximation (GGA-DFT). The full and dashed lines refer to the phonon dispersion corresponding to calculated unit cell parameters a=10.42(10.55) Å, b = 6.06 (6.17) Å, c = 4.75 (4.79) Å and a = 10.77 (10.66) Å, b = 6.20 (6.22) Å, c = 4.88 (4.83) Å for $LiFePO_4$ ($LiMnPO_4$). The zone-centre and zone-boundary phonon modes in $LiFePO_4$ and $LiMnPO_4$ soften at unit cell volume corresponding to the higher temperature. This region is hitherto defined by us as dynamically unstable regime.

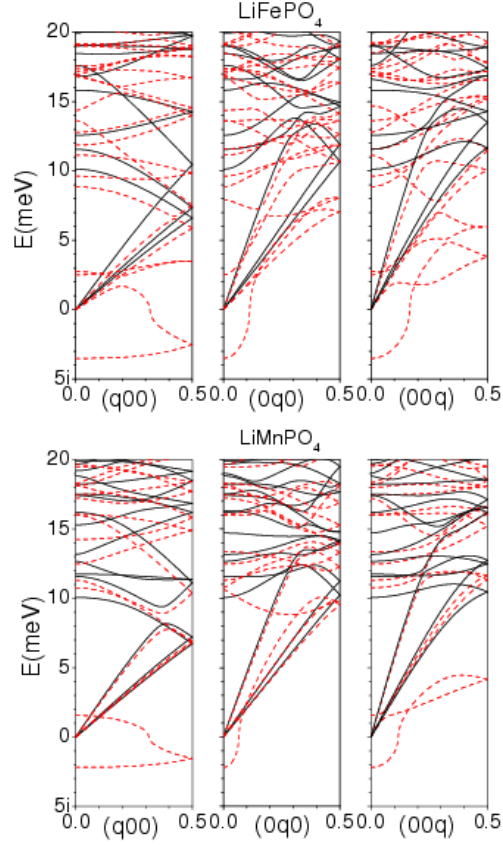

FIG. 8. (Color online) Zone-boundary and zone-centre modes as a function of unit cell volume. The zone-centre and zone-boundary phonon modes in $LiFePO_4$ and $LiMnPO_4$ soften at unit cell volume corresponding to higher temperatures.

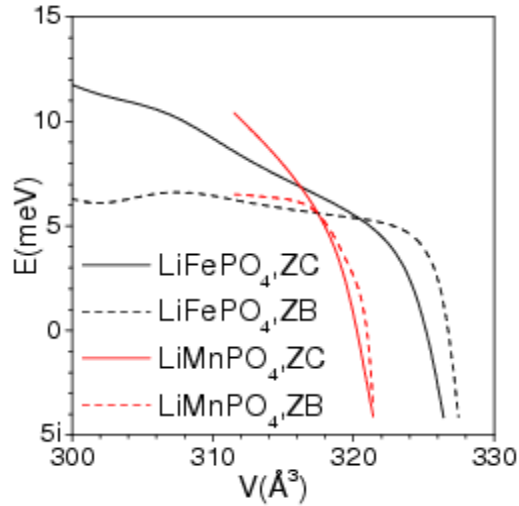

FIG. 9 (Color online) Motion of individual atoms for zone boundary and zone centre modes at unit cell volumes corresponding to ambient and dynamically unstable region. The numbers after the mode assignments give the phonon energies of mode in Fe(Mn) compound. *i* after the phonon energy indicates that mode is unstable. A 2×1×1 super cell of the primitive unit cell is shown for zone boundary mode at (0.5 0 0). The lengths of arrows are related to the displacements of the atoms. The zone-centre and zone-boundary phonon modes in LiFePO$_4$ and LiMnPO$_4$ soften at unit cell volume corresponding to higher temperatures. The absence of an arrow on an atom indicates that the atom is at rest. Key; Li: Red spheres, M=Mn or Fe: Yellow spheres, P: Green spheres, O: Blue spheres.

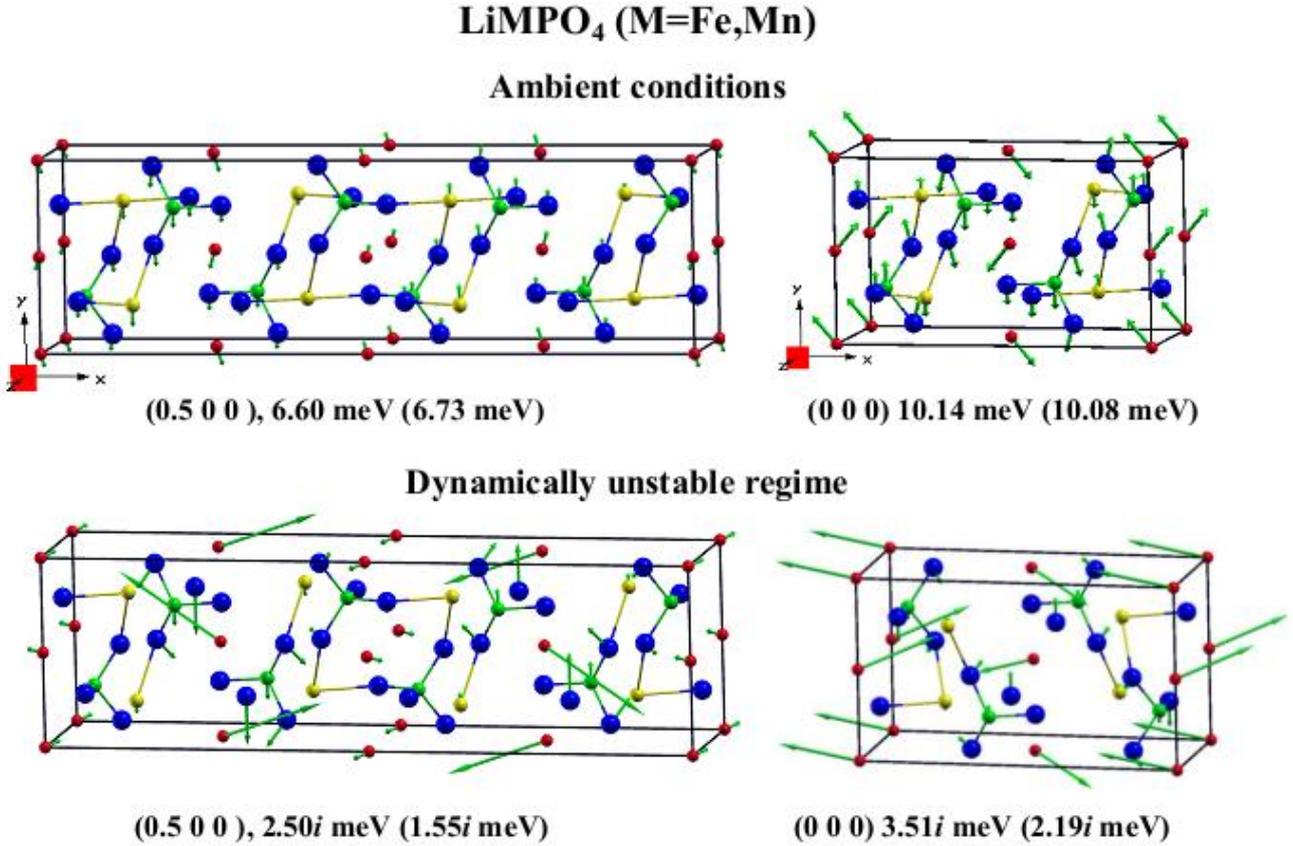

FIG. 10. (Color online) Comparison between the experimental [20,40] and calculated volume thermal expansion of LiMPO$_4$ (M=Mn, Fe). The dashed and full lines correspond to classical lattice dynamics (LD) and molecular dynamics (MD) calculations respectively.

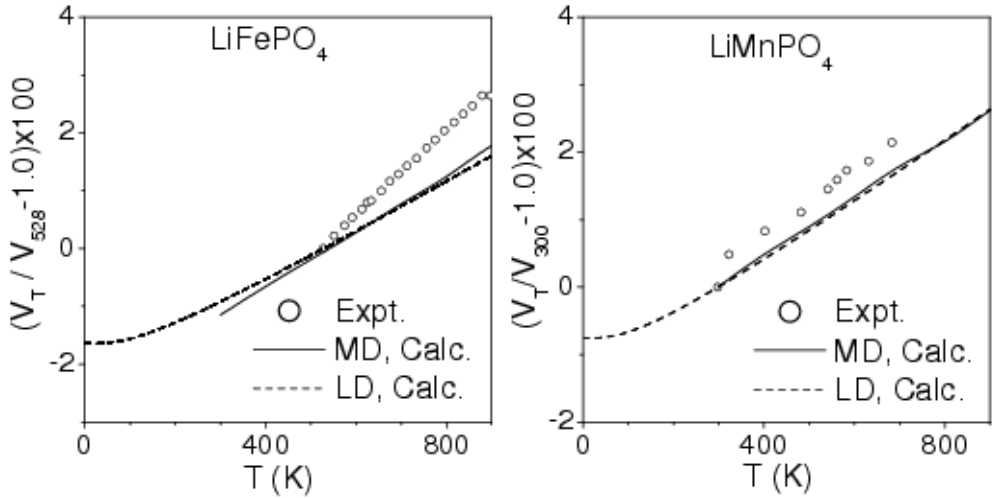

FIG.11. Mean-Squared Displacement of constituent atoms in LiFePO$_4$ and LiMnPO$_4$ with increasing temperature using molecular dynamics simulations.

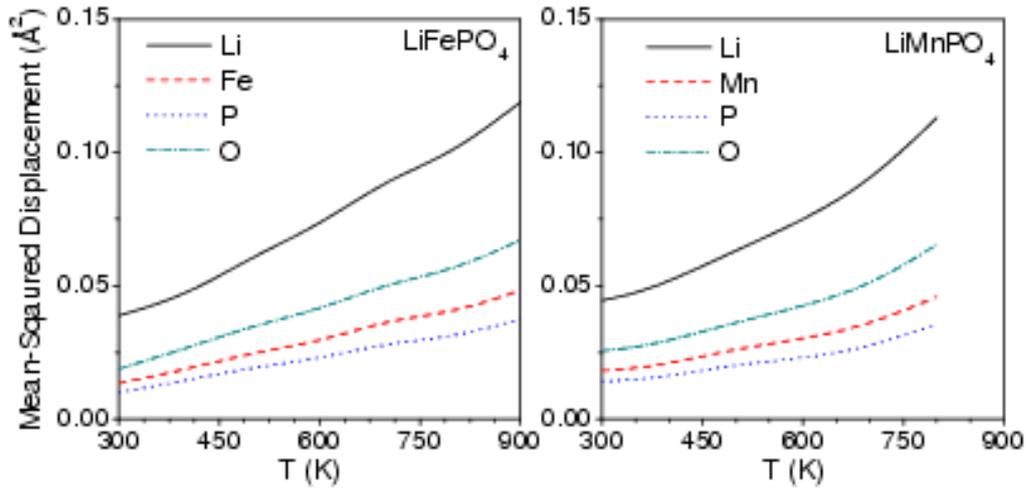

FIG. 12. Pair correlation function of different pairs of atoms in the olivines at various temperatures; 300K - Solid line ; 1100 K- dashed line.

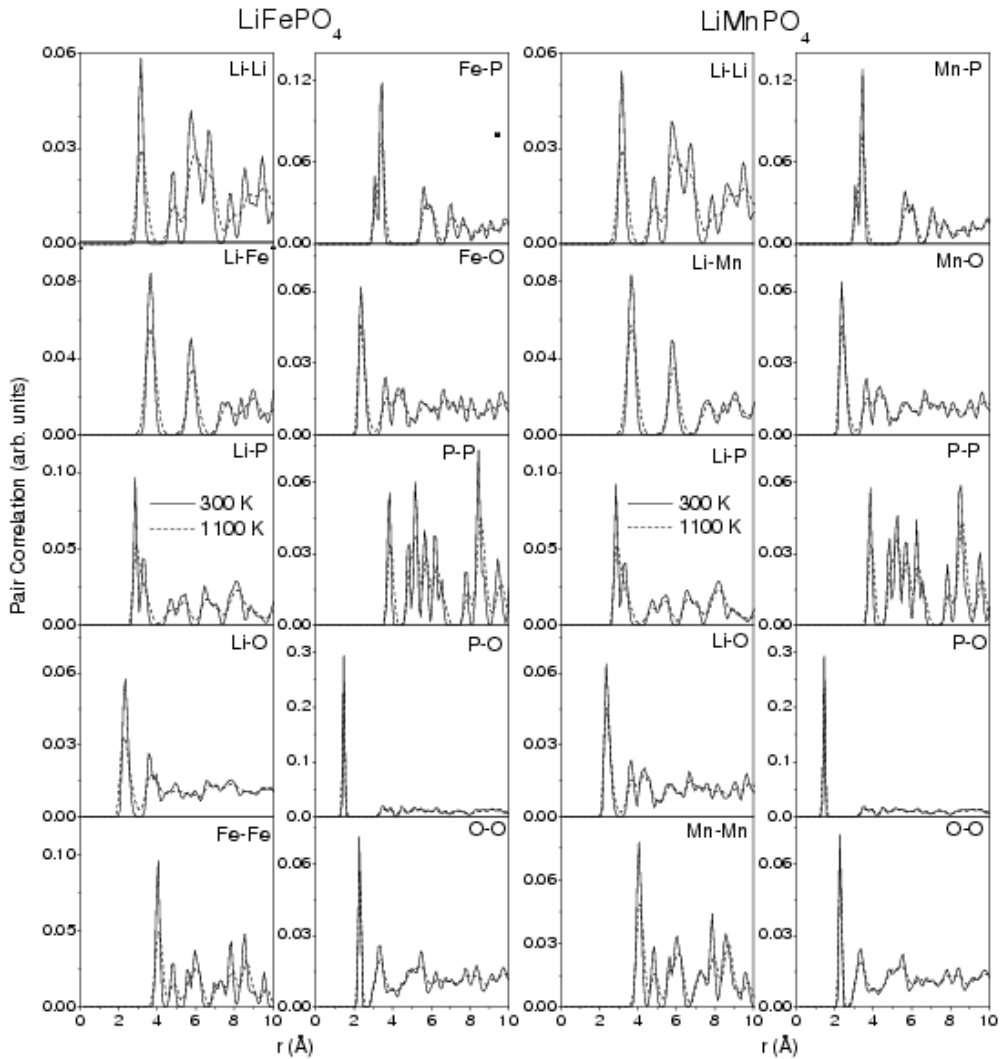